\documentclass[12pt,preprint]{aastex}

\usepackage{subfigure}
\usepackage{graphicx}
\usepackage{amsmath}
\usepackage{bbm}
\usepackage{bbold}
\usepackage{times}
\usepackage{natbib}
\usepackage{ulem}
\usepackage{color}
\newcommand{\mbf}{\boldsymbol} % make symbols bold in math mode
\newcommand{\I}{\mathrm{i}} % make complex i 

\slugcomment{Global helioseismology of meridional flow}

\shorttitle{Deeply penetrating meridional flow consisting of multiple flow cells}
\shortauthors{Schad et al.}

\begin{document}

\title{Global helioseismic evidence for a deeply penetrating Solar meridional flow consisting of multiple flow cells}

\author{A. Schad\altaffilmark{1,2}}
\affil{Kiepenheuer-Institut f\"ur Sonnenphysik, D-79104 Freiburg, Germany}\author{J. Timmer\altaffilmark{2,3}}
\affil{Institute of Physics, University of Freiburg, D-79104 Freiburg, Germany}
\and
\author{M. Roth}
\affil{Kiepenheuer-Institut f\"ur Sonnenphysik, D-79104 Freiburg, Germany}
\email{ariane.schad@kis.uni-freiburg.de}

\altaffiltext{1}{Institute of Physics, University of Freiburg, D-79104 Freiburg, Germany}
\altaffiltext{2}{Freiburg Center for Data Analysis and Modeling, University of Freiburg, D-79104 Freiburg, Germany}
\altaffiltext{3}{BIOSS Centre for Biological Signaling Studies, University of Freiburg, D-79104 Freiburg, Germany}

\begin{abstract}
We use a novel global helioseismic analysis method to infer the meridional flow in the deep Solar interior. The method is based on the perturbation of eigenfunctions of Solar $p$ modes due to meridional flow. We apply this method to time series obtained from Dopplergrams measured by the Michelson Doppler Imager aboard the Solar and Heliospheric Observatory (SOHO) covering the observation period 2004--2010. Our results show evidence that the meridional flow reaches down to the base of the convection zone. The flow profile has a complex spatial structure consisting of multiple flow cells distributed in depth and latitude. Toward the Solar surface, our results are in good agreement with flow measurements from local helioseismology.
\end{abstract}


\section{Introduction}
Most of our knowledge of the Sun's internal meridional circulation is provided by local helioseismic techniques such as ring-diagram analysis~\citep{hill88, haber02, gonzalez08} and time-distance helioseismology~\citep{duvall93, zhao04}. They provide reliable flow measurements down to $\approx20$\,Mm below the surface, where an on average poleward directed meridional flow is found with horizontal velocity amplitudes of about $20\,\mathrm{m\,s}^{-1}$ at mid-latitudes. The amplitude profile in the deeper interior, especially the depth of the return flow, remains uncertain, but these are critical quantities that determine the Solar cycle length in some dynamo models~\citep{dikpati99} and allow one to constrain hydrodynamic models of Solar convection~\citep{miesch12}. 

Inferences obtained from the advection of supergranules down to a depth of 70\,Mm \citep{hathaway12} and from apparent shifts of $p$-mode frequencies \citep{mitra-kraev07}, suggest shallow flow reversals within the upper convection zone. But generally it is assumed that the flow penetrates down to the base of the convection zone, although convincing evidence is still missing. Inferences from frequency shifts~\citep{braunfan98,krieger07} at deeper layers are especially in doubt~\citep{gough10}, since eigenfrequency perturbations are of second order in the flow~\citep{roth08}. However, the perturbation of $p$-mode eigenfunctions is of first order~\citep{woodard00,schad11,vorontsov11} and of central importance in deriving a global helioseismic meridional flow analysis method~\citep{schad11,schad12,woodard13}. 

In~\cite{schad11, schad12}, we suggested a new global helioseismic analysis method to infer the meridional flow in the deep interior by analyzing the perturbation of mode eigenfunctions of low and medium degree $l$. Our method was tested successfully on simulated data for different meridional flow models~\citep{schad11a,schad12}. The first applications to data from the Michelson Doppler Imager (MDI) instrument yielded promising measurements of the meridional flow component with the harmonic degree $s=2$ in the upper part of the Solar convection zone~\citep{schad12}.  

Here we present global helioseismic inferences of further harmonic components ($s=1$ to 8) and compare our results with flow measurements obtained from a local helioseismic analysis. We find global helioseismic evidence of a meridional flow that permeates the full convection zone and exhibits multiple flow cells in latitude and depth.

\section{Theory and Methods}
The meridional flow $\mbf{u}$ is modeled in terms of spherical harmonics $Y_{s}^{0}$ with the harmonic degree $s$ and azimuthal order $t=0$~\citep{roth08}
\begin{equation}
\label{eqn:flow-meridional}
\mbf{u}(r,\theta,\phi)=\sum_{s=1}^{\infty}\big[u_{s}(r)Y_{s}^{0}(\theta,\phi)\mbf{e}_{r}+v_{s}(r)\partial_{\theta}Y_{s}^{0}(\theta,\phi)\mbf{e}_{\theta} \big]\, .
\end{equation}
The degree $s$ specifies the number of flow cells over latitude, e.g., the $s=2$ component has two flow cells, one on each hemisphere. The spherical harmonic expansion coefficients $u_{s},v_{s}$ determine the radial and the horizontal flow strength with radius, respectively. They are, assuming mass-conservation, related to each other by~\citep{lavely92} 
\begin{eqnarray}
\label{eqn:uvrelation}
\rho_{0}r s(s+1)v_{s}=\partial_{r}(r^{2}\rho_{0}u_{s})\, ,
\end{eqnarray}
with $\rho_{0}$ as the mass density.
\subsection{Coupling of $p$ Modes}
The meridional flow leads to a coupling of $p$ modes which distorts the modes. The perturbed eigenfunction $\mbf{\xi}_{k}$ of a mode $k$ can be approximated in quasi-degenerate perturbation theory by~\citep{lavely92} 
\begin{equation}
\label{eq:modecontrib}
\mbf{\xi}_{k}(r, \theta, \phi)=\mbf{\xi}^{0}_{k}(r,\theta,\phi)+\sum_{k'\in K_{k}\setminus\{k\}} c_{kk'} \mbf{\xi}^{0}_{k'}(r,\theta,\phi)\, ,
\end{equation}  
where $\mbf{\xi}^{0}_{k'}$ is the eigenfunction of mode $k'$ of a flow-free Solar reference model. The triple $k=(n,l,m)$ refers to the radial order $n$, the harmonic degree $l$, and the azimuthal order $m$. The expansion coefficients $\{c_{kk'}\}$ are imaginary; their moduli represent the coupling strengths which contribute only to a subset $K_{k}\setminus\{k\}$ of modes adjacent to mode $k$~\citep{lavely92,roth08,schad11}. As the meridional flow is axial-symmetric, the coefficients $c_{kk'}$ possess azimuthal symmetry. They can be expanded in terms of suitable orthogonal polynomials $\{\mathcal{P}_{l'l}^{s}(m)\}$~\citep{schad11},
\begin{equation}
\label{eq:Hpolynom}
c_{kk'}=c_{nl,n'l'}(m)\approx \I\,\frac{\omega_{nl}}{\omega^{2}_{nl}-\omega^{2}_{n'l'}}\sum_{s}b_{n'l',nl}^{s}\mathcal{P}_{l'l}^{s}(m)\, ,
\end{equation}
where $\omega_{nl}$ and $\omega_{n'l'}$ are the unperturbed angular {frequencies} of the coupling modes. The expansion coefficients, denoted as $b$-coefficients, are related to the radial flow coefficients $u_{s}$ by a linear integral equation~\citep{schad11}
\begin{eqnarray}
\label{eqn:bcoeff}
b_{n'l',nl}^{s}=\int_{0}^{R}\rho_{0}(r)K^{n'l',nl}_{s}(r)u_{s}(r)r^{2}dr\, ,
\end{eqnarray}
where the poloidal flow kernel $K^{n'l',nl}_{s}(r)$ determines the sensitivity of the coupling modes on the meridional flow at radius $r$~\citep{lavely92}. 

\subsection{Observational Effect of Mode Coupling}
The coupling of modes manifests in a crosstalk between the observable global oscillations. 
To measure this effect we introduced the amplitude ratio, $y_{lm\,l'm}(\omega_{nlm}):=\frac{\tilde{o}_{l'm}(\omega_{nlm})}{\tilde{o}_{lm}(\omega_{nlm})}$,  between the Fourier amplitude of two spherical harmonic transformed global oscillations $\tilde{o}_{lm}$ and $\tilde{o}_{l'm}$ evaluated at the mode frequency $\omega_{nlm}$~\citep{schad11}. In the first order, the amplitude ratio is related to the coupling coefficients by~\citep{schad11} 
\begin{align}
\label{eq:aratio}
y_{lm\,l'm}(\omega_{nlm})\approx\frac{\sum_{k''\in K_{k}} c_{kk''}L_{k'k''} \xi^{r}_{k''}(R)}{\sum_{k''\in K_{k}} c_{kk''}L_{kk''}\xi^{r}_{k''}(R)} \in \mathbb{C}\, ,
\end{align}
where $c_{kk}=1$ and $\xi^{r}_{k''}(R)$ is the radial eigenfunction of mode $k''$ at the observation point $R$. The matrix elements $\{L_{k'k''}\}$ denote the systematic leakage of spectral power of a mode $k$ to nearby modes $k'$, since spherical harmonics are not orthogonal when not integrated over the full sphere~\citep{schou94,korzennik04}. 
The amplitude ratio represents a random quantity. Its expectation value $G_{lm,l'm}(\omega_{nlm}):=\langle y_{lm,l'm}(\omega_{nlm})\rangle$ is related to the cross-spectrum \mbox{$CS_{lm,l'm}=\langle{\tilde{o}^*_{lm}}\tilde{o}_{l'm}\rangle$} by 
\begin{align}
\label{eq:cgaindef}
G_{lm,l'm}(\omega_{nlm})=\frac{CS_{lm,l'm}(\omega_{nlm})}{S_{lm}(\omega_{nlm})}\, ,
\end{align}
where $S_{lm}$ is the auto-spectrum of $o_{lm}$ and $()^*$ denotes the complex conjugate. We refer to $G_{lm,l'm}(\omega_{nlm})$ as \textit{complex gain} in analogy to the gain in filter theory~\citep{Oppenheim75}. It can be estimated from finite time series by replacing the spectrum and cross-spectrum in Equation~\eqref{eq:cgaindef} with appropriately chosen sample estimators. The variance $\sigma^{2}$ of the sample estimator of $G_{lm,l'm}(\omega)$ can be estimated by~\citep{Hannan70}
\begin{align}
\label{eqn:estgainvar}
\sigma^{2}(\omega)=\frac{1}{dof-2}\frac{S_{l'm}(\omega)}{S_{lm}(\omega)}\left(1-Coh^{2}(\omega)\right)\, ,
\end{align}
where $Coh$ is the coherency spectrum of $o_{lm}$ and $o_{l'm}$~\citep{schad08} and the parameter $dof$ defines the equivalent number of degrees of freedom~\citep{Hannan70} of the spectral estimator used to estimate $CS_{lm,l'm}$, $S_{lm}$, $S_{l'm}$, and $Coh$. 

\section{Data Analysis and Results}
\subsection{Data Analysis}
We investigate time series of global oscillations obtained from the medium-$l$ structure program of the MDI instrument~\citep{scherrer95} covering the years 2004--2010. The data are analyzed in six blocks, each with a length of approximately 360 days, in order to compensate for possible periodic annual variations of the systematic spatial leakage, e.g., due to variations in the $B_{0}$-angle~\citep{zaatri06}.

We evaluate modes of harmonic degrees $0\leq l\leq 198$ and frequencies $1.32\,\mathrm{mHz}\leq \nu_{nl}\leq 4.77\,\mathrm{mHz}$. For each mode and data block, the amplitude ratios and errors are estimated using Equation~\eqref{eq:cgaindef} and \eqref{eqn:estgainvar}. In total, 12925 pairs of coupling multiplets $(k=(n,l),k'=(n',l'))$ with harmonic separation $dl=|l'-l| \leq 10$ and frequency separation $|\nu_{k}-\nu_{k'}|\leq \delta \nu= 31\,\mu$Hz are analyzed. The amplitude ratios are averaged over the six blocks and symmetrized with respect to azimuthal order $m$ in order to reduce the estimation errors and to remove effects from differential rotation, respectively. The $b$-coefficients are estimated from the amplitude ratios on the basis of Equations~\eqref{eq:Hpolynom} and \eqref{eq:aratio} by means of a standard non-linear least-squares fitting routine and the leakage matrix of the MDI instrument. The horizontal component of the leakage matrix is neglected in the analysis, since its contribution to Equation~\eqref{eq:aratio} is small compared to the radial component for modes of low and medium degree $l$. The 
radial flow coefficients of degree $s=1,\,\hdots,8$ are estimated from the $b$-coefficients, see Equation~\eqref{eqn:bcoeff}, by means of the standard inversion technique of Subtractive Optimally Localized Averages (SOLA,~\citep{pijpers94}). The poloidal flow kernels are computed from the eigenfunctions of Solar Model S~\citep{dalsgaard96}. 

The inversion is carried out on a grid of target positions $\{r_{j}\}_{j=1,\hdots,J}$ in the range $0.57 < r_{j}/R < 0.992$, where $R$ is the Solar radius. For each degree $s$, the regularization parameters entering the SOLA inversion are adjusted to obtain flow estimates as deep as possible on the one hand and well localized inversion kernels on the other. The radial position and resolution of flow estimates is determined by the center of mass and standard deviation of the Gaussian shaped inversion kernels, respectively~\citep{dalsgaard90}. 

The horizontal flow coefficients $v_s(r)$ are estimated from the radial flow coefficients $u_s(r)$ using equation~\eqref{eqn:uvrelation} by fitting polynomials of degree $N=3$ to the term $r^{2}\rho_0 u_{s}(r)$ on intervals of grid length $\Delta_{q}=19$ centered around each target position to derive the radial derivatives. The error of the radial flow coefficients is derived from the Hessian matrix of the least-squares fit and the error propagation of the SOLA inversion analysis, where we take into account the non-uniform estimation errors of the amplitude ratios. The errors of the horizontal flow coefficients are estimated by means of a resampling approach in order to take into account the influence of correlations between flow estimates at adjacent target positions by the polynomial fit.   

\subsection{Inversion Results for Single Flow Components}
\begin{figure}[t]
\begin{center}
\includegraphics[width=15cm]{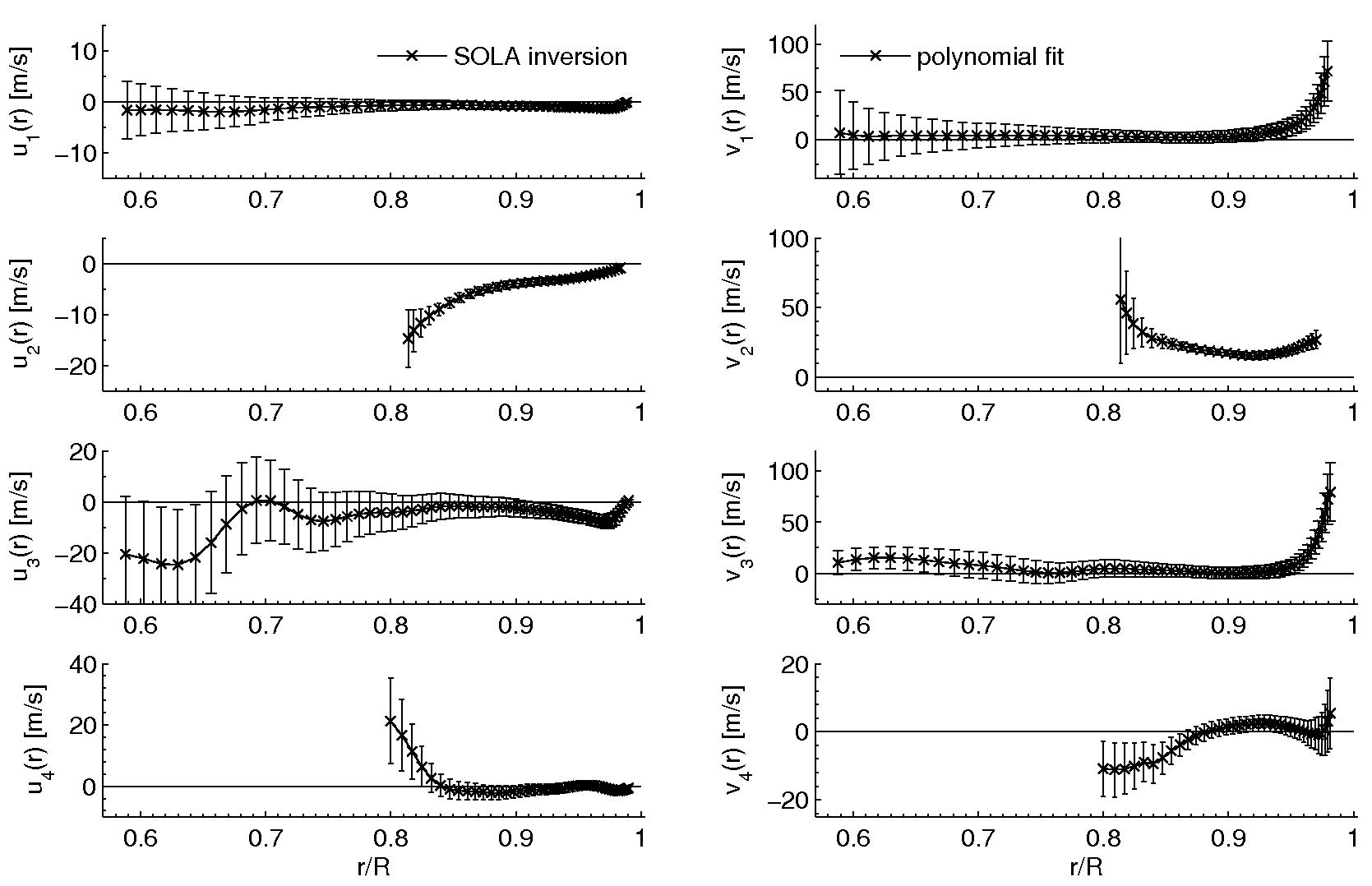}
\caption{Radial flow coefficients $u_{s}$ (left) and horizontal flow coefficients $v_{s}$ (right) of the meridional flow as a function of radius $r/R$ for the degrees $s=1,\hdots,4$ with the $1\sigma$-error estimated from MDI data using the global analysis approach. \label{fig1}}
\end{center}
\end{figure}

\begin{figure}[t]
\begin{center}
\includegraphics[width=15cm]{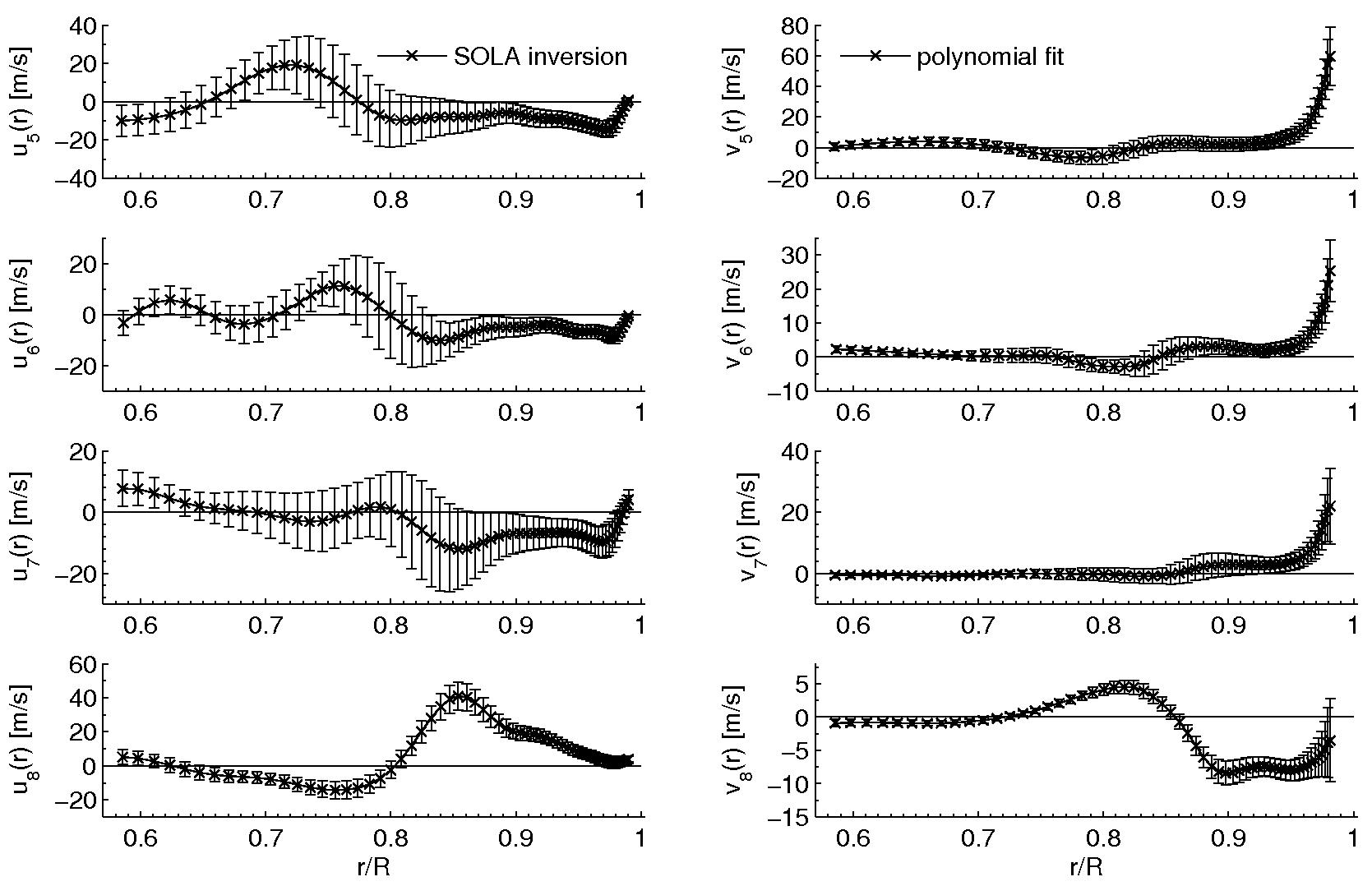}
\caption{Estimated flow coefficients $u_{s}$ (left) and $v_{s}$ (right) as presented in Figure~\ref{fig1}, but for the degrees $s=5,\hdots,8$. \label{fig1b}}
\end{center}
\end{figure}

The estimated radial and horizontal flow coefficients ${u}_{s}(r)$ and ${v}_{s}(r)$ are presented in Figure~\ref{fig1} for the degrees $s=1,\hdots,4$ and in Figure~\ref{fig1b} for the degrees $s=5,\hdots,8$ as a function of radius. All radial flow coefficients fade toward the surface, which is physically meaningful as the flow is expected to be confined inside the Sun. We obtain significant flow components for the degrees $s=2$ and $s=8$. 

The $s=2$ component has a slow radial flow coefficient $u_{2}(r)$ which increases approximately linearly with depth and vanishes toward the surface. The horizontal flow coefficient $v_{2}$ is approximately constant over radii that cover $(0.82\pm 0.02)\leq r/R\leq (0.970\pm 0.009)$. 

For the $s=8$ flow we find that the horizontal component is approximately constant in most of the outer convection zone while the corresponding radial component increases linearly with depth. A zero crossing of $u_{8}$ at $r=(0.80 \pm 0.02)$ indicates the presence of two layers of flow cells in depth. One layer is located in the upper 2/3 of the convection zone and a second layer of weaker counterflow cells is located in the lower part of the convection zone and might extend even deeper. The horizontal flow coefficient $v_{8}$ is significant down to $r=(0.72\pm 0.03)\,R$ where it vanishes. It shows a flow reversal at $r=(0.86\pm 0.02)\,R$.

For $s=4$, we do not observe a significant flow coefficient in most of the convection zone. For $r<0.84\,R$, the results indicate the top of a layer of flow cells. However, further data are needed to confirm this observation. For $s=6$, we do not observe significant flow coefficients at all, except near the surface. For the flow components of odd degrees $s$, we find well localized inversion kernels throughout the whole convection zone and even below. However, for $ r\lesssim 0.9\,R$, the estimated flow coefficients are not significant. Compared to the even degrees, the errors of the flow estimates of adjacent odd degrees $s$ are,  in general, larger. This can be attributed to weak or non-existent mode couplings. Due to the decreased coherency, the estimated amplitude ratios have large errors which propagate through the data inversion approach, see Equation~\eqref{eqn:estgainvar}.

For the components $s=1,3,5,6,7$, we observe a significant systematic drop of the radial flow coefficient at $0.9<r/R<1$ to negative values. This results in an unreliable large horizontal flow coefficient in the Solar subsurface that is mediated by the polynomial fit approach. The actual origin of this drop remains unclear. We assume a systematic bias which might result from the non-linear estimation of $b$-coefficients when carried out on insignificant mode couplings. This will be subject of future investigations. 

\subsection{Reconstruction of the Total Flow}
\begin{figure}[t]
\begin{center}
\includegraphics[width=13cm]{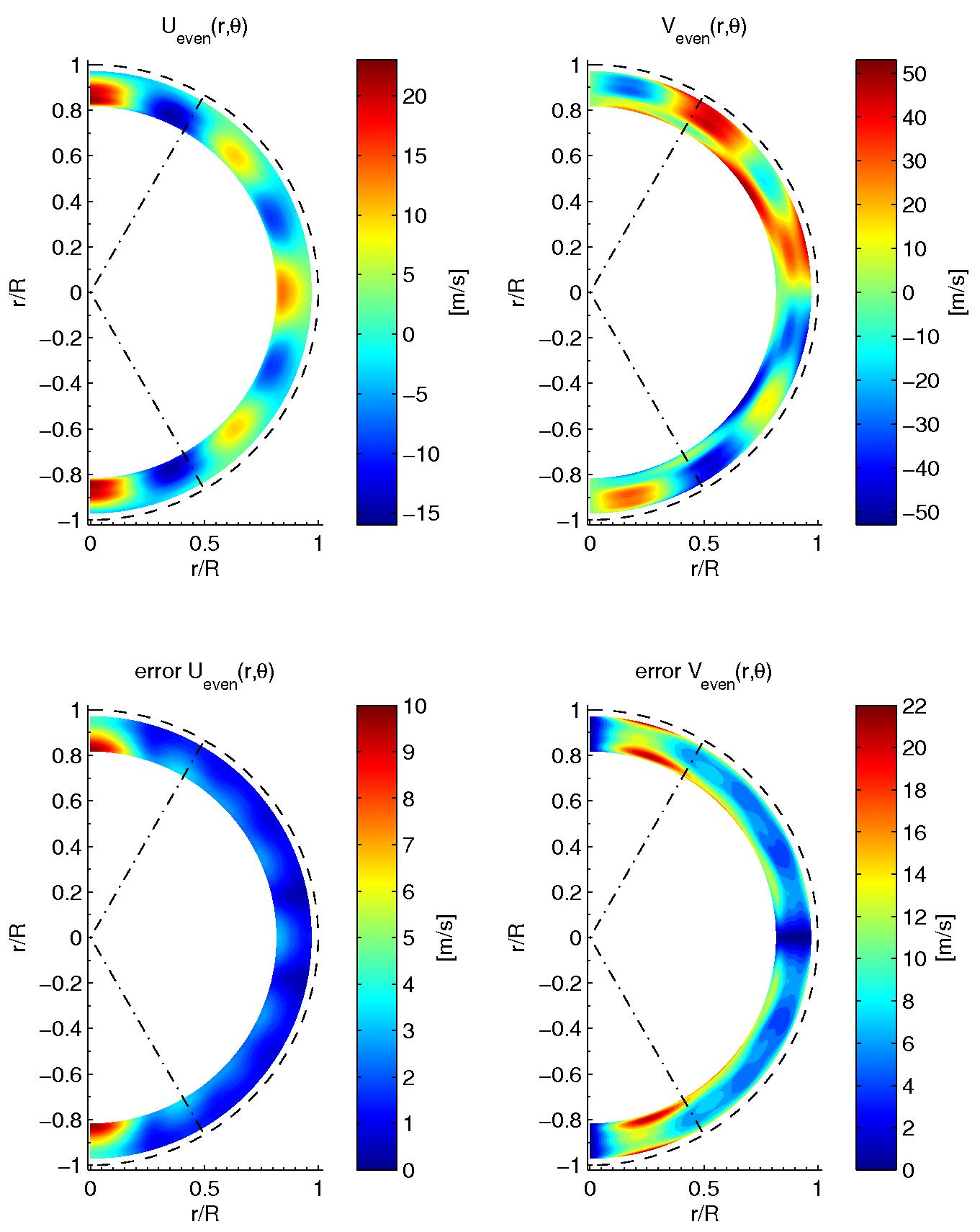}
\caption{Top: Cross-sections through the composite meridional flow summed over even degrees $s=2,4,6,8$ in the $(r,\theta)$-plane between $0.82\leq r/R\leq 0.97$. The dashed line marks the Solar surface. Dashed-dotted lines mark the latitudes $\theta=\pm 60^{\circ}$. Left: Radial flow $U_{even}$; positive (negative) values correspond to outward (inward) directed flows. Right: Horizontal flow $V_{even}$; positive (negative) values correspond to northward (southward) directed flows. Bottom: Standard error of the composite flow. Left: 1$\sigma$-error of $U_{even}$. Right: 1$\sigma$-error of $V_{even}$.\label{fig4}} 
\end{center}
\end{figure}

The total radial and horizontal flow, $U,V$, are computed by summation over the single flow components $U_{s}(r,\theta)=u_{s}(r)Y_{s}^{0}(\theta,\phi)$ and $V_{s}(r,\theta)=v_{s}(r)\partial_{\theta}Y_{s}^{0}(\theta,\phi)$ according to Equation~\eqref{eqn:flow-meridional}. Since the odd components do not indicate a significant flow, we restrict the summation to even degrees $s=2,4,6,8$. Cross-sections of the flow profiles for $U_{even}$ and $V_{even}$ and the respective 1$\sigma$-errors are depicted in Figure~\ref{fig4}. Note the total flow is presented only for the region where all even components overlap, i.e. $0.82\leq r/R\leq 0.97$,  although the $s=8$ component has a significant amplitude throughout the whole convection zone.

The total meridional flow reveals a complex spatial pattern. Below the subsurface layer and in the upper part of the convection zone, the flow is dominated by the $s=8$ flow component. At larger depths, the $s=2$ component becomes dominant showing a poleward directed flow on both hemispheres. The estimation error typically increases with depth. The speed of the composite horizontal flow at mid latitudes $\theta=\pm45^{\circ}$ is about $(20\pm 6)\,\mathrm{m\,s}^{-1}$ over a large range of radii $0.83\lesssim r/R\lesssim 0.96$. At high latitudes, $|\theta|>60^{\circ}$, the results  must be interpreted carefully, since there the observation of Solar oscillations is limited due to projection effects. 

\subsection{Comparison with Subsurface Flow Measurements}
In a ring-diagram analysis of GONG data from 2001 July to 2002 August by \cite{komm05} the meridional flow was measured between 0.6 and16\,Mm, and was separated into a large- and a residual small-scale component with respect to latitude by fitting polynomials.

The large-scale flow component corresponds to a superposition of the $s=2$ and $s=4$ components in our flow representation, the residual flow represents a superposition of the remaining components. We compare these measurements with the significant results obtained from our analysis, namely the $s=2$ and $s=8$ flow components. 

\begin{figure}[t]
\begin{center}
\includegraphics[width=13cm]{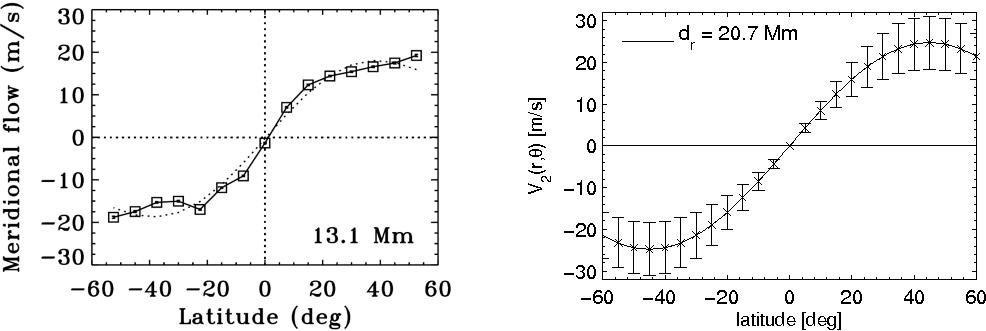}
\caption{Left: Horizontal meridional velocity $V(r,\theta)$ at 13.1\,Mm depth as a function of latitude obtained from a ring-diagram analysis of GONG data by~\cite{komm05}. The dotted line represents a large-scale flow component obtained from a polynomial fit. Right: Horizontal flow component $V_{2}(r,\theta)$ at $20.7$\,Mm depth with $1\sigma$-error obtained from our global inversion approach. Positive (negative) values correspond to northward (southward) directed flows. \label{fig2}} 
\end{center}
\end{figure}

The large-scale horizontal flow is directed poleward and increases with depth in accordance with our measurements. The latitudinal profile obtained at 13.1\,Mm by \cite{komm05} is compared with our global helioseismic measurement of $V_{2}(r,\theta)$ at $(20.7\pm6)$\,Mm in Figure~\ref{fig2}. The profiles and amplitudes are similar although measured at slightly different depths. 

\begin{figure}
\begin{center}
\includegraphics[width=15cm]{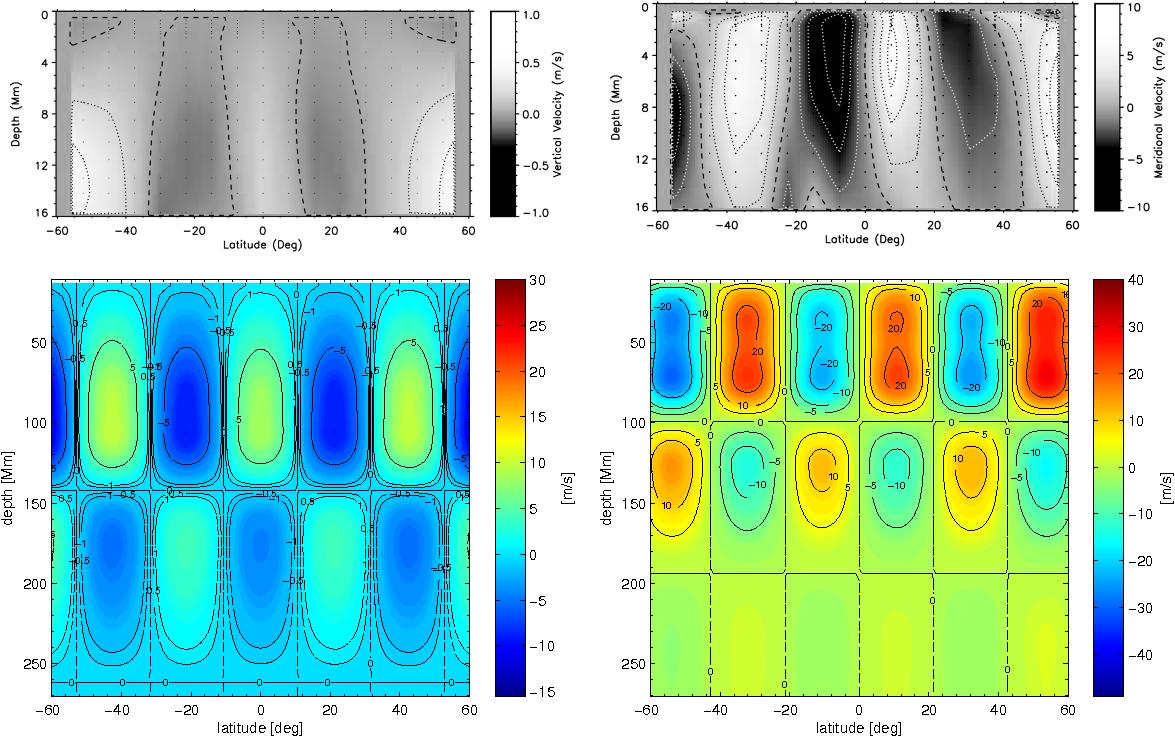}
\caption{Top: Residual meridional radial velocity $U_{res}$ (left) and residual horizontal velocity $V_{res}$ (right) between 0.6--16\,Mm depth and $\pm60^\circ$ latitude obtained from ring-diagram analysis~\citep{komm05}. Dashed lines indicate contour lines of zero velocity. Bottom: The radial flow component $U_{8}$ (left) and the horizontal flow component $V_{8}$ (right) between 13--270\,Mm depth and $\pm60^\circ$ latitude. Isolines of selected velocity for $U_{8}=\pm 0,0.1,0.5,1\,\mathrm{m\,s}^{-1}$ and $V_{8}=\pm 0,5,10,20\,\mathrm{m\,s}^{-1}$ are highlighted by black lines. Left: Positive (negative) values correspond to radially outward (inward) directed flows. Right: Positive (negative) values correspond to northward (southward) directed flows.\label{fig3}}
\end{center}
\end{figure}

Maps of the small-scale radial and horizontal components between a depth of 0.6--16\,Mm from ring-diagram analysis are shown in the top of Figure~\ref{fig3}, our global flow measurements for the $s=8$ component obtained at a depth of $(13\pm 4)$--$(270\pm25)$\,Mm are shown at the bottom. The residual flow exhibits several flow cells over latitude. We find similarity between both measurements concerning the number and orientation of the flow cells with respect to latitude. Within the depth range of 13--16\,Mm, where both measurements cover the same depths, the flows are of the same magnitude: $|U_{res}|,|U_{8}| \lesssim 1\,\mathrm{m\,s}^{-1}$ and $|V_{res}|,|V_{8}| \lesssim 10\,\mathrm{m\,s}^{-1}$. In addition, the horizontal flow measurement $V_{8}$ shows flow reversals at depths of $(100\pm 13)$ and $(195\pm 20$)\,Mm, reflecting two layers of counterflow cells.

\section{Discussion and Conclusions}
We have measured the meridional flow in the deep Solar interior from MDI data covering 2004--2010 by means of a new global helioseismic analysis method introduced previously. The method uses perturbations of $p$-mode eigenfunctions of low and medium degrees to probe both the horizontal and radial component of the meridional flow in the deep interior. 

The results do not provide a complete picture of the internal total flow, but we find significant amplitudes for the component of degree $s=2$ down to 0.82\,R directed poleward, and for the $s=8$ component down to the base of the convection zone with flow reversals. The results presented here provide global helioseismic evidence of a spatially complex structured meridional flow which consists of multiple flow cells distributed in latitude and depth, and which reaches down at least to the base of the convection zone. Such a meridional flow profile is also favored by hydrodynamic simulation studies of the convection zone~\citep{kueker11,miesch12}.

The approach used here is complementary to local helioseismic techniques using frequency shifts or perturbations of wave travel times to measure the internal velocity field. Near the Solar surface, our flow measurements are consistent with measurements from ring-diagram analyses of GONG data covering 2001--2009~\citep{komm05,gonzalez2010}, which show a residual small-scale modulation of the meridional flow in the Solar subsurface. This modulation is discussed as a meridional component of the torsional oscillation of the zonal flow that is present throughout the convection zone~\citep{vorontsov02,howe05}. However, local helioseismic measurements were not able to probe this small-scale meridional flow component beyond the subsurface layer. We find such a modulation in our global helioseismic analysis as well, and identify it with a flow component of harmonic degree $s=8$. Our results reveal that this subsurface modulation represents only the head of a layer of flow cells which extends throughout the whole convection zone. Therefore, we suspect a global coupling mechanism between the zonal and the meridional flow residuals.

Typically, flux-transport dynamo models assume a large-scale $s=2$ meridional circulation~\citep{wang91, choudhuri95,kueker01, charbonneau10}. The relevance of the multi-cellular flow presented here, especially the role of the layer of small-scale weaker counter-flow cells in the lower part of the convection zone, for the Solar dynamo action need to be investigated from time-resolved meridional flow measurements.

Very recently, when this work was under review, \cite{zhao13} presented measurements of the horizontal meridional flow from a time-distance analysis of data from the Helioseismic and Magnetic Imager (HMI,~\citep{scherrer12,schou12}) recorded after 2010. They also found a complex flow profile reaching deep into the convection zone but with a less complex structure in latitude. The differences might be due to the analysis of different record lengths and observation periods. They compensated for a systematic artifact which is suspected to originate from the center-to-limb-variation of the HMI spectral absorption line~\citep{zhao12}. As discussed by~\cite{woodard13}, such a variation might also affect analyses of mode eigenfunction perturbations and could result in an overestimation of the flow amplitude. However, their eigenfunction perturbation analysis differs from ours, especially in the observables used and the theoretical assumptions made for modes of low degree. Additionally, the effect on the MDI absorption line is likely smaller~\citep{zhao13} and, compared to time-distance analysis, our method weights data differently toward the Solar limb. Our current considerations suggest that such an effect might be small in our analysis, especially for modes of degree $l<100$.

The results presented here are obtained using Solar Model S and the leakage matrix of the MDI instrument. A refinement of both, e.g., taking into account center-to-limb variations affecting the observation of global oscillations and other atmospheric influences on the eigenfunctions, might further improve meridional flow estimates. The analysis of such systematic effects will be subject of future studies.

Furthermore, in future investigations we will evaluate further mode couplings to infer the flow components $s=2$ and $s=4$ at depths below $0.82\,R$, and to completely determine the profile of the total meridional flow. We also aim to improve the sensitivity of the flow measurements by extending our analysis to data from MDI for 1996--2011 and from the HMI available since 2010. In addition, the higher resolution of HMI data should improve inferences of the meridional flow at high latitudes.  

\acknowledgments
This work utilizes data from SOHO/MDI. We thank J. Schou for providing the
MDI leakage matrix, T. Larson for providing the global oscillation time series, and R. Komm, G. R\"udiger, and R. Arlt for useful discussions. The research leading to these results has received funding from the European Research Council under the European Union's Seventh Framework Program (FP/2007-2013) / ERC Grant Agreement n. 307117 and from the Deutsche Forschungsgemeinschaft DFG, Grant Ti 315 4/2.

%\bibliographystyle{apj}
%\clearpage

\clearpage

%% FIGURES:
%% Fig1

%% Fig1b

%% Fig2

%% Fig3

%% Fig4


\begin{thebibliography}{}

\bibitem[Braun \& Fan(1998)]{braunfan98} Braun, D. C., Fan, Y. 1998, ApJL, 508, L105

\bibitem[Charbonneau(2010)]{charbonneau10} Charbonneau, P. 2010, LRSP, 7, 3

\bibitem[Choudhuri et al.(1995)]{choudhuri95} Choudhuri, A.R., Sch\"ussler, M., Dikpati, M. 1995, A\&A, 303, L29

\bibitem[Christensen-Dalsgaard et al.(1996)]{dalsgaard96} Christensen-Dalsgaard, J., D\"appen, W., Ajukov, S. V., et al. 1996, Sci, 272, 1286

\bibitem[Christensen-Dalsgaard et al.(1990)]{dalsgaard90} Christensen-Dalsgaard, J., Schou, J., \& Thompson, M. J. 1990, MNRAS, 242, 353

\bibitem[Dikpati \& Charbonneau(1999)]{dikpati99} Dikpati, M., Charbonneau, P. 1999, ApJ, 518, 508

\bibitem[Duvall et al.(1993)]{duvall93} Duvall, T. L., Jr., Jefferies, S. M., Harvey, J. W., \& Pomerantz, M. A. 1993, Natur, 362, 430

\bibitem[Gonz\'alez Hern\'andez(2008)]{gonzalez08} Gonz\'alez Hern\'andez I. 2008, JPhCS, 118, 012034

\bibitem[Gonz\'alez Hern\'andez et al.(2010)]{gonzalez2010} Gonz\'alez Hern\'andez, I., Howe, R., Komm, R., \& Hill, F. 2010, ApJL, 713, L16

\bibitem[Gough \& Hindman(2010)]{gough10} Gough, D., \& Hindman, B.W. 2010, ApJ 714, 960

\bibitem[Haber et al.(2002)]{haber02}Haber, D. A., Hindman, B.W., Toomre, J., et al. 2002, ApJ, 570, 855

\bibitem[Hannan(1970)]{Hannan70} Hannan, E. J. 1970, Multiple Time Series (New York: Wiley)

\bibitem[Hathaway(2012)]{hathaway12} Hathaway, D. 2012, ApJ,760, 84 

\bibitem[Hill(1988)]{hill88} Hill, F. 1988, ApJ, 333, 996

\bibitem[Howe et al.(2005)]{howe05} Howe, R., Christensen-Dalsgaard, J., Hill, F., et al. 2005, ApJ, 634, 1405

\bibitem[Komm et al.(2005)]{komm05} Komm, R., Howe, R., Hill, F., Gonz\'alez Hern\'andez, I., \& Toner, C. 2005, ApJ, 631, 636

\bibitem[Korzennik et al.(2004)]{korzennik04} Korzennik, S. G., Rabello-Soares, M. C., \& Schou, J. 2004, ApJ, 602, 481

\bibitem[Krieger et al.(2007)]{krieger07} Krieger, L., Roth, M., v.d. L\"uhe, O. 2007, AN, 328, 252

\bibitem[K\"uker et al.(2001)]{kueker01} K\"uker, M., R\"udiger, G., \& Schultz, M. 2001, A\&A, 374, 301

\bibitem[K\"uker et al.(2011)]{kueker11} K\"uker, M., R\"udiger, G., Kitchatinov, L.L. 2011, A\&A, 530, A48

\bibitem[Lavely \& Ritzwoller(1992)]{lavely92} Lavely, E. M. \& Ritzwoller, M. H. 1992, RSPTA, 339, 431

\bibitem[Miesch et al.(2012)]{miesch12} Miesch, M. S., Featherstone, N. A., Rempel, M., \& Trampedach, R. 2012, ApJ, 757, 128

\bibitem[Mitra-Kraev \& Thompson(2007)]{mitra-kraev07} Mitra-Kraev, U., \& Thompson, M. J. 2007, AN, 328, 1009

\bibitem[Oppenheim \& Schafer(1975)]{Oppenheim75} Oppenheim, A. V. \& Schafer, R. W. 1975, Digital Signal Processing (Englewood Cliffs, NJ: Prentice-Hall)

\bibitem[Pijpers \& Thompson(1994)]{pijpers94} Pijpers, F. P. \& Thompson, M. J. 1994, A\&A, 281, 231

\bibitem[Roth \& Stix(2008)]{roth08} Roth, M., \& Stix, M. 2008, SoPh, 251, 77

\bibitem[Schad et al.(2008)]{schad08} Schad, A., Roth, M., Schelter, B., von der L\"uhe, O., \& Timmer, J. 2008, in JPhCS, 118, 012091

\bibitem[Schad et al.(2011a)]{schad11} Schad, A., Timmer, J., \& Roth, M. 2011a, ApJ, 734, 97 

\bibitem[Schad et al.(2011b)]{schad11a} Schad, A., Roth, M., \& Timmer, J. 2011b, in JPhCS, 271, 012079

\bibitem[Schad et al.(2012)]{schad12} Schad, A., Timmer, J., \& Roth, M. 2012, AN, 333, 991

\bibitem[Scherrer et al.(1995)]{scherrer95} Scherrer, P. H., Bogart, R. S., Bush, R. I., et al. 1995, SoPh, 162, 129

\bibitem[Scherrer et al.(2012)]{scherrer12} Scherrer, P. H., Schou, J., Bush, R. I., et al. 2012, SoPh, 275, 207

\bibitem[Schou \& Brown(1994)]{schou94} Schou, J. \& Brown, T. M. 1994, A\&AS, 107, 541

\bibitem[Schou et al.(2012)]{schou12} Schou, J., Scherrer, P. H., Bush, R. I., et al. 2012, SoPh, 275, 229

\bibitem[Vorontsov(2011)]{vorontsov11} Vorontsov, S. V. 2011, MNRAS, 418, 1146

\bibitem[Vorontsov et al.(2002)]{vorontsov02} Vorontsov, S. V., Christensen-Dalsgaard, J., Schou, J., Strakhov, V. N., \& Thompson, M. J. 2002, Sci, 296, 101

\bibitem[Wang \& Sheeley (1991)]{wang91} Wang, Y.-M., \& Sheeley, N. R., Jr. 1991, ApJ, 375, 761

\bibitem[Woodard(2000)]{woodard00} Woodard, M. F. 2000, SoPh, 197, 11

\bibitem[Woodard et al.(2013)]{woodard13} Woodard, M. F., Schou, J., Birch, A. C., \& Larson, T. P. 2013, SoPh, 287, 129

\bibitem[Zaatri et al.(2006)]{zaatri06} Zaatri, A., Komm, R., Gonz\'alez Hern\'andez I., Howe, R., \& Corbard, T. 2006, SoPh, 236, 227

\bibitem[Zhao et al.(2004)]{zhao04} Zhao, J., Kosovichev, A. G., \& Duvall, T. L., Jr. 2004, ApJ, 607, L135

\bibitem[Zhao et al.(2012)]{zhao12} Zhao, J., Nagashima, K., Bogart, R. S., Kosovichev, A. G., \& Duvall, T. L., Jr. 2012, ApJ, 749, L5

\bibitem[Zhao et al.(2013)]{zhao13} Zhao, J., Bogart, R. S., Kosovichev, A. G., \& Duvall, T. L., Jr., \& Hartlep, T. 2013, ApJL, 774, L29 
  
\end{thebibliography}
\end{document}